\documentclass[%
 reprint,
unsortedaddress,
nofootinbib,
 amsmath,amssymb,
 aps,
 prx,
floatfix,
]{revtex4-2}
\makeatletter
\renewcommand\frontmatter@abstractwidth{\dimexpr\textwidth\relax}

\makeatother
\usepackage{comment}
\usepackage{graphicx}
\usepackage{dcolumn}
\usepackage{bm}
\usepackage[dvipsnames]{xcolor}
\usepackage{mathrsfs}

\usepackage{graphicx}
\usepackage{dcolumn}
\usepackage{bm}
\usepackage[utf8]{inputenc}
\usepackage[T1]{fontenc}
\usepackage{mathptmx}
\usepackage{etoolbox}
\usepackage{amsmath}
\usepackage{caption}
\usepackage{svg}
\usepackage{float}
\usepackage[normalem]{ulem}
\usepackage[dvipsnames]{xcolor}
\usepackage{amssymb}

\begin{document}

\preprint{APS/123-QED}

\title{Local Rules for Directing the Emergence of Global Properties in Complex Structures}

\author{Andrew Slezak, Varda F. Hagh}
\email[]{hagh@illinois.edu}
\affiliation{%
Mechanical Science and Engineering, University of Illinois at Urbana-Champaign, Urbana, Illinois 61801, USA}%

\begin{abstract}

From ants to caterpillars, many biological systems composed of simple builders have been observed to construct complex, adaptive, and functional architectures without requiring complete access to the global state of the structure. In these systems, global function emerges from the accumulation of local actions, as individual builders follow local rules to manipulate, modify, and deposit material in response to local environmental stimuli. This raises the question of how local rules can be selected for simple builders so that desired functions reliably emerge as a natural consequence of their interactions with their environment. We propose a systematic framework for determining such rules and demonstrate its effectiveness using a minimal model inspired by tent caterpillars and their silk networks. Using our framework, we show that local rules can be designed so that when simple builders follow them during network construction, the values of several emergent properties including area coverage, mean line density, and front curvature can be directed toward specific target values. We use a statistical approach to determine how rules can be modified to increase the probability that a useful local change occurs, the magnitude of that change, or both, so that the target property can be achieved reliably. Our results demonstrate a general strategy for linking local rules to emergent global properties in complex structures. This strategy offers a step toward fabricating functional structures using simple builders in uncertain environments where global information, precise control, and sustained human supervision are infeasible.  

\end{abstract}

\maketitle

\section{Introduction}

\begin{figure}[h!]
    \centering
    \includegraphics[width=0.9\linewidth]{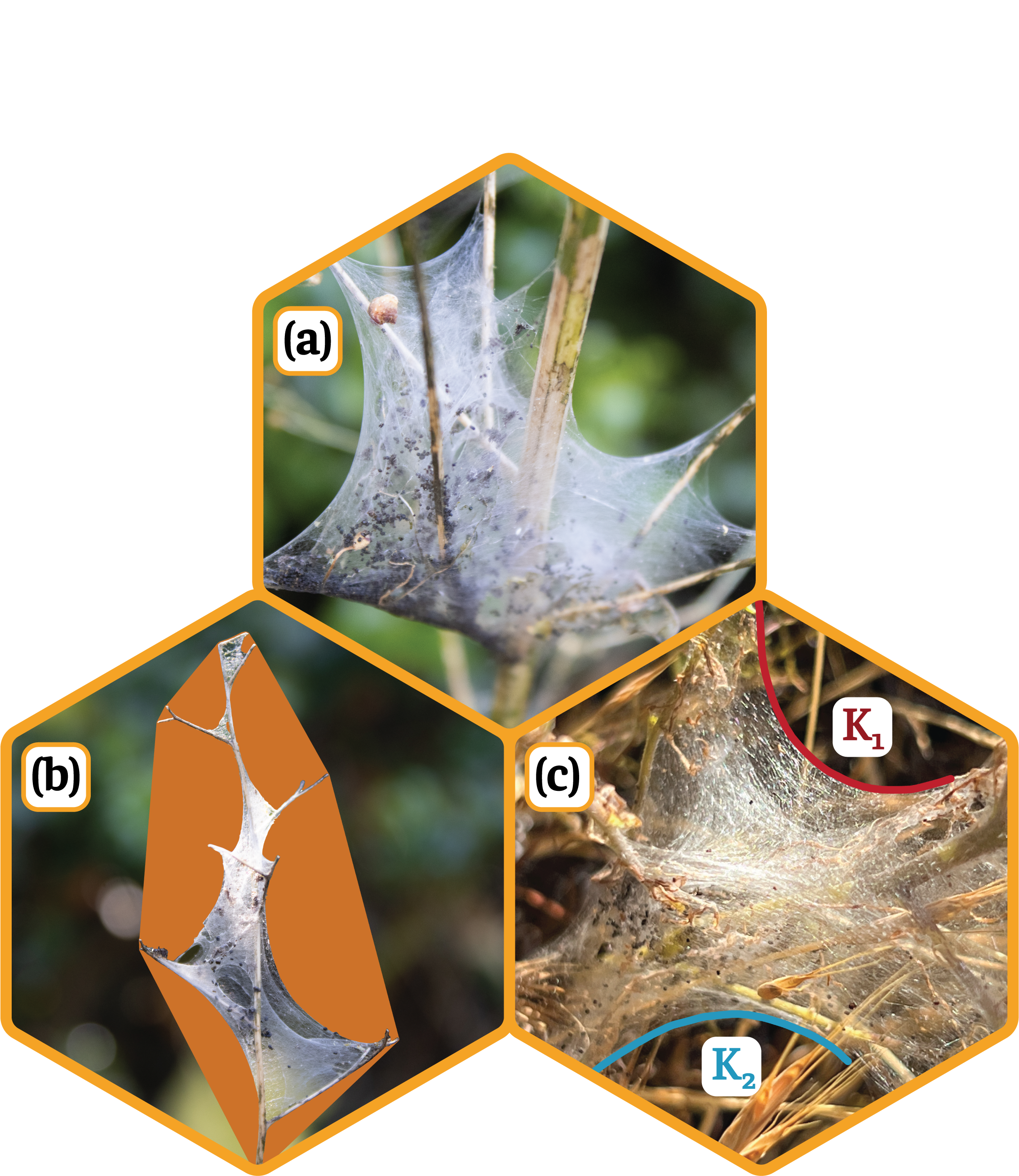}
    \caption{\small{Communal nests (tents) built by Chalcedon Checkerspot caterpillars (\textit{Euphydryas chalcedona}). (a) The tents are network structures made of silk fibers that adapt to irregular boundary conditions in the environment. (b) The tents occupy only a fraction of the maximum available area or volume shown by the orange outlined region. (c) Multiple distinct curvatures can emerge on the surface of the tents during construction. Photos taken by Varda F. Hagh \& Avaneesh Narla at Jasper Ridge Biological Preserve.}}
    \label{fig:hexagonal_images1}
\end{figure}

Throughout nature we observe many striking examples of how complex functional architectures can be built without centralized design. For instance, many arthropods including spiders, net-spinning caddisflies, glow-worms, bees, ants, termites, and tent caterpillars, all build intricate structures whose diverse forms are specifically adapted to local environmental conditions and support the functional needs of their species~\cite{caine_architecture_2026, jackson_insect_2026, riiska_physics_2024, fard_crystallography_2022, peleg_collective_2018, su_imaging_2018, hansell_animal_2005, wiggins_caddisflies_2005}. These systems are especially remarkable because their builders operate without top-down planning, specialized tools, or the explicit calculations that guide engineered systems. Yet they reliably produce structures like those shown in Fig.~\ref{fig:hexagonal_images1}, which have been observed to regulate temperature, retain humidity in arid environments, and remain mechanically robust during storms~\cite{fitzgerald_tent_1995, joos_roles_1988, fitzgerald_tent-building_1983}. Such global functional properties emerge not from an explicit blueprint, but from the ways in which organisms manipulate, deposit, and organize material building blocks.

In many of these systems, the local organization of material building blocks during construction is not arbitrary. Builders often follow \textit{local rules}, defined here as behavioral patterns by which they select, modify, deposit, or connect material building blocks in response to their immediate environment. Through repeated application, these local rules shape the growth of the structure and influence its overall properties. The link between such local rules and the emerging global properties in the structures point to a broader design principle in which simple builders, referred to here as agents, can construct functional structures without centralized control. This capability would be invaluable in environments such as disaster zones, the deep ocean, or distant planetary surfaces, where precise control and sustained human oversight are not feasible. 

Although recent advances in autonomous control and swarm robotics have shown that simple agents can assemble complex structures by following case-specific behavioral rules~\cite{zhu_self-organizing_2024, talamali_when_2021, petersen_review_2019, soleymani_bio-inspired_2015, napp_distributed_2014, giardina_robotectonics_2026, liu_design_2025, werfel_designing_2014, werfel_collective_2007, theraulaz_coordination_1995}, a general methodology for designing local rules that produce structures with prescribed global properties remains entirely lacking. Thus we propose a systematic method to develop local rules that direct selected properties to reliably emerge in the absence of traditional centralized control schemes. However, a central challenge is that the relationship between local rules and emergent global properties in complex structures is generally nonlinear, stochastic, and many-to-many. A given set of rules can often produce many diverging outcomes as noise propagates through the construction process. This complexity makes it difficult to design local rules that can \textit{reliably} yield prescribed outcomes, despite how prolific this appears to be in nature.

In this work, we formulate the design of local construction rules as an inverse design problem. Given a desired global property, the goal is to determine how simple builders (\textit{i.e.}, agents), should manipulate material building blocks such that the target property emerges in the completed structure. We introduce a framework for solving this inverse problem that builds on concepts from \textit{tunable matter}, where macroscopic properties in materials can be controlled by tuning their microscopic degrees of freedom~\cite{goodrich2015principle, rocks2017designing, hagh2018disordered, hagh2019broader, pashine2019directed, rocks2019limits, hexner2020periodic, hagh2022transient, kim2022structural, hexner2023training, stern2023learning, rouviere2023emergence, dillavou_machine_2024, bhaumik2024mechanical, stern2025physical, arzash2025rigidity, guo2026learning}. Here, these degrees of freedom correspond to the properties of building blocks and their interactions with neighboring blocks or other materials in the environment.

To accomplish this, we take inspiration from tent caterpillars, which build with silk fibers to produce network structures on their host plants. We study a minimal model of these caterpillars in which our agents construct networks by depositing line segments within branch-like scaffolds. Here the deposited line segments are the fundamental building blocks of the structure. In this case, their spatial placement and length are the primary tunable degrees of freedom, whose values determine the resulting network architecture and shape its geometric properties. However, without prior knowledge, we cannot always know apriori what properties are trivial or intractable to direct. 

Therefore we first establish a fully stochastic baseline in which agents assign random values to these degrees of freedom. This baseline tests whether a global property of interest can emerge without developing more complex rules, and identifies the range of values accessible through unbiased sampling. If needed we then introduce modifications to rules which either bias the process by increasing the probability of a useful local update occurring, or by increasing the magnitude a given update will impart on the property of interest, or both.
Using this approach, we demonstrate that simple local rules can be designed to reliably direct global properties such as area coverage, mean line density, and front curvature in the resulting networks. By \textit{directing} these properties, we mean that the ensemble mean approaches a specified target while variability across independent realizations remains sufficiently small. Although we focus here on geometric properties of $2D$ networks, the conceptual framework is not restricted to these listed observables and can be extended to other physical properties and to $3D$ systems.

\begin{figure}[h!]
    \centering
    \includegraphics[width=1\linewidth]{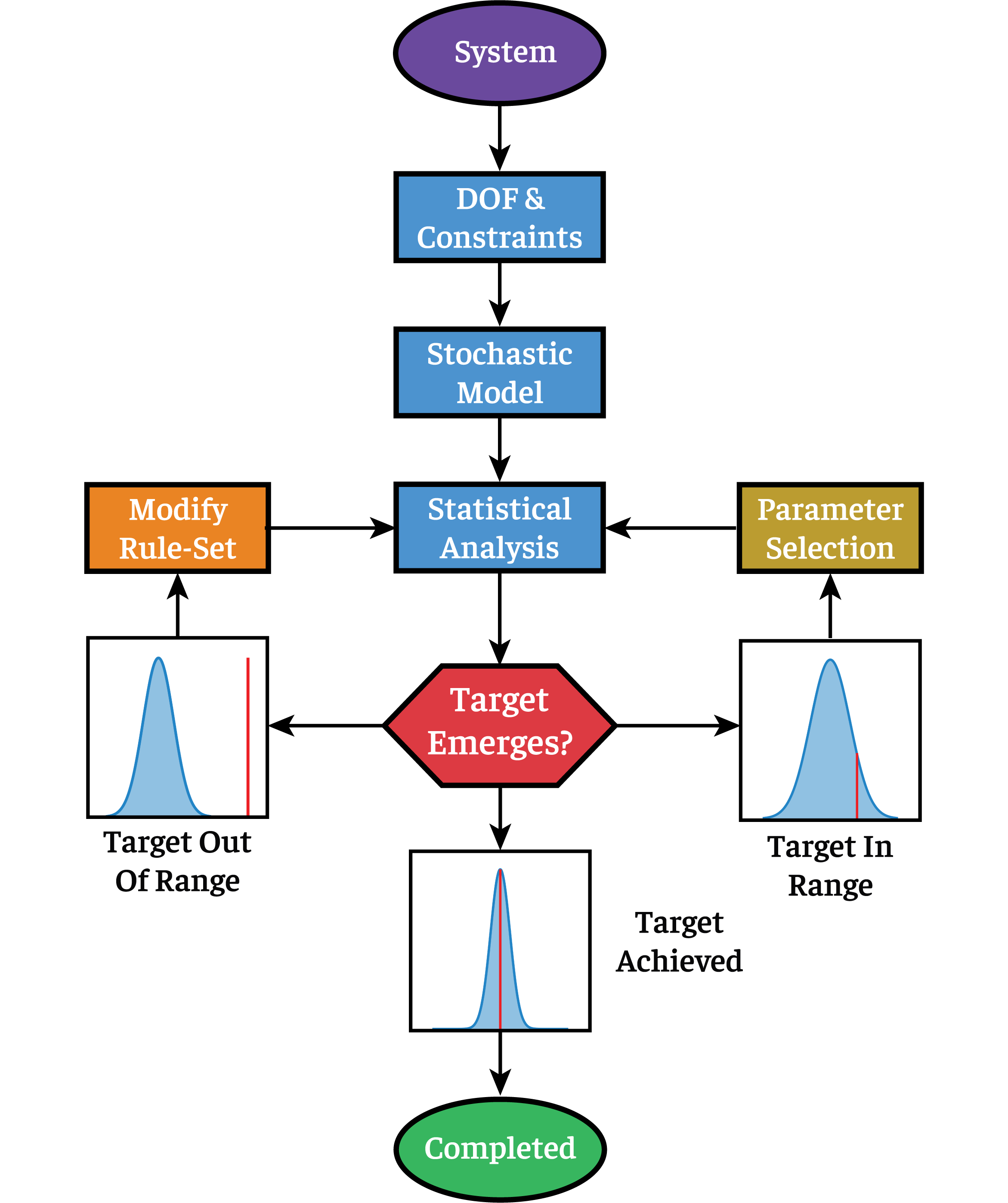}
    \caption{\small{Schematic of the methodology for designing local rules. The system and its building blocks define the accessible degrees of freedom (DOF) and constraints. Assigning random values to the DOF within the boundaries imposed by constraints produces a stochastic baseline used to identify whether the target property emerges reliably and what range of values is accessible. If the target lies within this range, parameter values are selected to reach it. Otherwise, the rule-set is modified and the procedure is repeated.}}
    
    \label{fig:rule_design}
\end{figure}

\section{Methodology}

To address the many challenges inherent to this task, we begin by specifying the system and its building blocks. For a given property of interest (surface area, shape, compliance, thermal capacity, \textit{etc}.), we determine which accessible degrees of freedom can be tuned to modify that property. In the present work, we focus on the geometric properties of network structures. The building blocks are therefore line segments, whose spatial positions and lengths serve as the tunable degrees of freedom in the system. The set of tunable degrees of freedom remains fixed throughout the process, while local rules determine how their values are selected or modified within the available range imposed by the boundary conditions.

Because the space of possible local rules is vast, we first establish a stochastic baseline in which values are assigned to the tunable degrees of freedom with a uniform random probability. This baseline provides a neutral reference case that reveals which values of the property are accessible without any directed rule design. For the network structures studied here, this corresponds to randomly sampling line segment positions and lengths within the accessible region. Directed rules are then introduced by replacing random sampling with state-dependent choices that bias the evolution of the system toward a desired outcome.

To formalize this idea, we consider a construction process in which an observable quantity $Q$, associated with an evolving structure, changes through a sequence of local updates. Here, $Q$ represents any global property of interest, such as area coverage or density. As a minimal description, we focus on additive processes in which each local update adds a new building block that either increases $Q$ or leaves it unchanged. For a specified target value $Q^{*} $, we define a useful update as one that changes $Q$ in the direction of the target. At time step $t$, such an update changes $Q$ by an amount $\delta Q_{\rm t}$. The expected value of change in $Q$ at time step $t$, denoted by $\langle \Delta Q\rangle_t$, is then
\begin{equation}
\langle \Delta Q \rangle_{\rm t} = \sum p_{\rm t} \delta Q_{\rm t},
\label{eq:expected-value}
\end{equation}
where $p_{\rm t}$ is the probability that a useful modification occurs at time $t$, $\delta Q_{\rm t}$ is the corresponding change in $Q$, and the sum is taken over all possible useful updates. A rule can direct $Q$ toward $Q^{*}$ by increasing the probability of useful updates, increasing the change produced by each useful update, or both. In the agent-based implementation described below, these mechanisms correspond, respectively, to decision rules that determine whether an agent builds or moves, and to build rules that determine how the local degrees of freedom are tuned.

To increase $\delta Q_{\rm t}$, a rule can restrict the values assigned to the tunable degrees of freedom to a subset that is expected to produce a larger useful change, rather than sampling uniformly from all available values. We will later demonstrate this mechanism in the area coverage section with a maximum distance constraint. Changing $p_{\rm t}$ requires a rule that affects when useful actions occur, such as movement or building. Since these actions are governed by the decision rule, changes to $p_{\rm t}$ must act through that rule.
A general strategy we employ is to allow agents to estimate a local measure of the target observable $Q$, compare that estimate with a prescribed threshold, and use the outcome to decide whether to build or move. 

We note that this formalism only describes the expected change in the observable $Q$, but does not capture fluctuations around that expectation. To evaluate whether a rule \textit{reliably} directs a given property to emerge, we must also account for variability across independent realizations. For a given tuning rule $\mathfrak{R}$, an ensemble of $m$ independent realizations produces not a single value of $Q$, but a distribution of values $Q_i$ where $i=1, 2, \dotsc,m$. From this distribution, we can compute an ensemble average $\overline{Q}(\mathfrak{R})$ and its associated standard deviation $\sigma_{\rm Q}$.

\begin{figure}[h]
    \centering
    \includegraphics[width=1\linewidth]{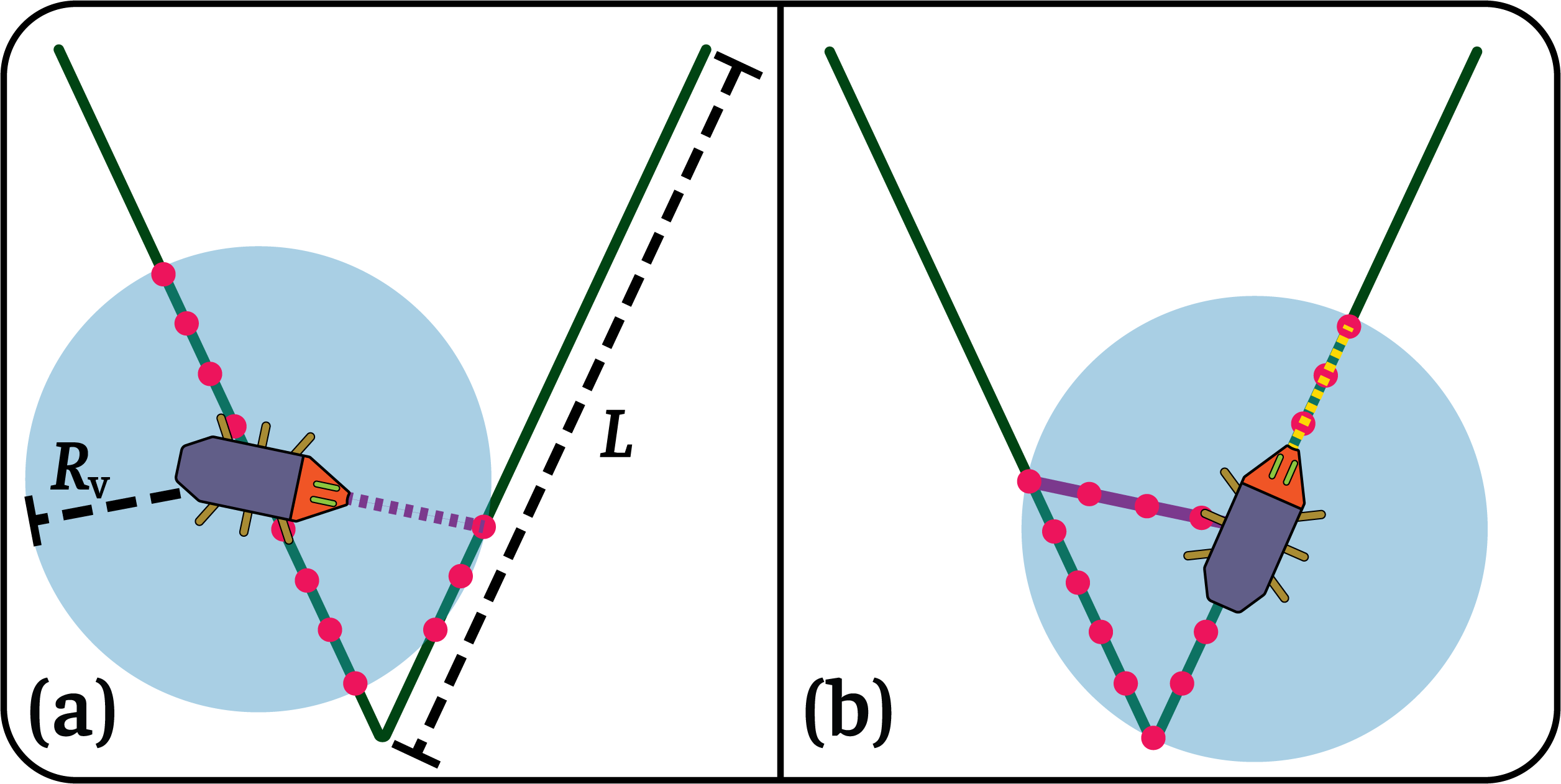}
    \caption{\small{(a) The agent begins the time step by discretizing the environment within its radius of vision into a set of points. It then selects a point from this list based on the rule-set it follows, constructs a line to that point, and repositions to the endpoint of the line. (b) In the following time step, the agent again discretizes its environment, including the newly deposited line, and repeats the process.}}

    \label{fig:agent_construction}
\end{figure}

We define success in directing an observable to a target value in terms of both accuracy and precision. Specifically, the ensemble average $\overline{Q}(\mathfrak{R})$ must lie within an acceptable tolerance of the target value, while the standard deviation must remain sufficiently small. Thus, the objective is not simply that $Q \rightarrow Q^{*}$, but rather
\begin{equation}
    |Q^{*} - \overline{Q}(\mathfrak{R})| \leq \epsilon \qquad \& \qquad \sigma_{\rm Q} \leq \sigma^*_{\rm Q},
\end{equation}
where $\epsilon$ and $\sigma^*_{\rm Q}$ define acceptable tolerances for accuracy and precision, respectively.

Once a rule has been observed to promote the emergence of the property of interest, the value of the property can be directed toward a specific target by adjusting the rule's input parameters. These parameters do not change the structure of the rule but rather determine where the system operates within the range of outcomes enabled by the rule.  Generally, a target value $Q^{*}$ is considered accessible when it lies within the range of values produced by a rule and its associated parameters. In our agent-based model, the primary parameters are the radius of vision, $R_{\rm v}$, the number of agents, $N_{\rm a}$, and time $t$, although additional parameters may be introduced depending on the target property. The radius of vision sets the spatial range over which an agent can sense (\textit{i.e.}, measure) and act, thereby controlling the locality of both information acquisition and construction. The number of agents determines how many local updates can be attempted in parallel, while time affects the total number of local updates permitted for each agent. Thus, rule modification and parameter selection are distinct steps: first, a directed rule is designed to bias $p_{\rm t}$, $\delta Q_{\rm t}$, or both. Then, parameter values such as $R_{\rm v}$, $N_{\rm a}$, or $t$ are selected to realize a desired outcome. This procedure is illustrated schematically in Fig.~\ref{fig:rule_design}.

We group rules followed by the agents into behavioral \textit{rule-sets}. Each rule-set consists of three components: a decision rule, a build rule, and a movement rule. The decision rule determines whether an agent builds or moves at a given time step. The build rule specifies which segment is deposited when construction occurs, and the movement rule specifies where the agent relocates when construction does not occur. This decomposition allows us to distinguish rules that primarily affect the probability of useful updates, $p_{\rm t}$, from those that primarily affect the expected change in the property of interest produced by a construction event, $\delta Q_{\rm t}$.

\begin{figure*}
    \centering
    \includegraphics[width=1\linewidth]{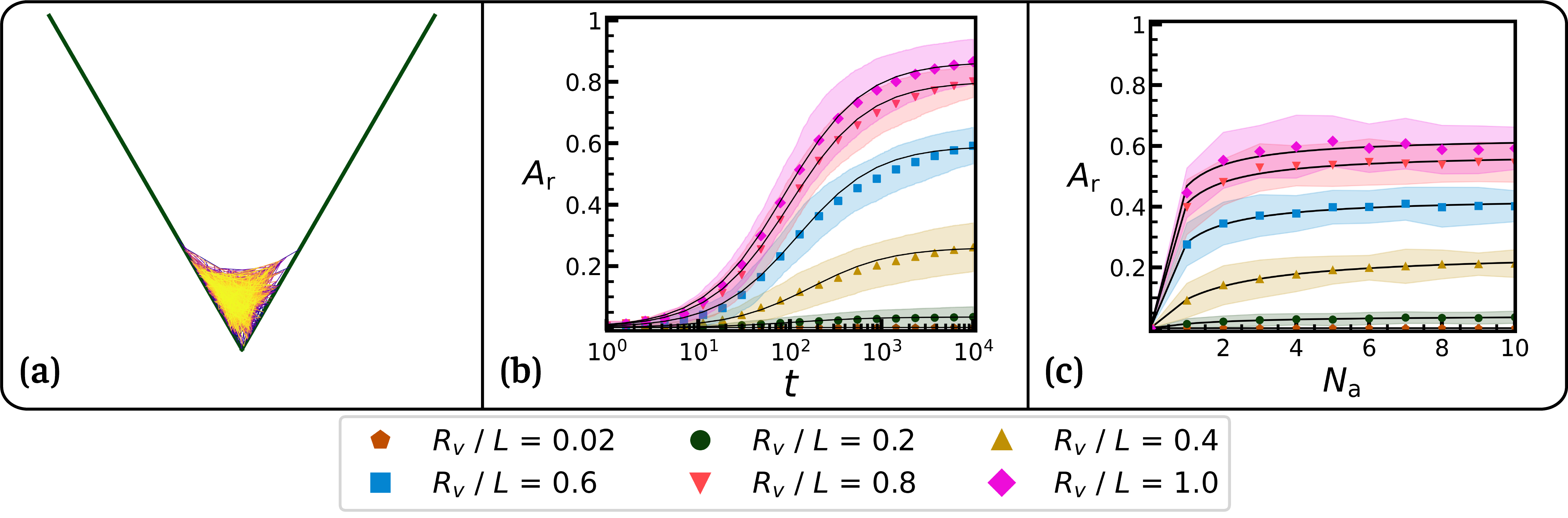}
    \caption{\small{(a) Example network formed by a single agent following the stochastic baseline rule-set after $t=10^4$ time steps with $R_{\rm v}/L=0.2$. (b) Area coverage ratio, $A_{\rm r}$, as a function of time, $t$, for various values of $R_{\rm v}/L$ under the stochastic baseline rule-set. Black solid lines show fits to Eq.~\ref{eq:saturation}. (c) Area coverage ratio, $A_{\rm r}$, as a function of the number of agents, $N_{\rm a}$, after $t=100$ time steps for various values of $R_{\rm v}/L$ under the stochastic baseline rule-set. Black solid lines show fits to Eq.~\ref{eq:saturation}. Shaded regions indicate one standard deviation across $100$ independent simulations.}}
    \label{fig:null-area}
\end{figure*}

Here, we model agents as point-like active particles that move within their environment and deposit line segments from points within it; in our case this environment consists of the scaffold and the growing network structure. At each time step, an agent senses the region within its radius of vision, $R_{\rm v}$, and uses measurements available within that local region to decide whether to build or move. A build action deposits a line segment from the agent's current position to a selected endpoint and then updates the agent's position to that endpoint, as shown in Fig.~\ref{fig:agent_construction}. In contrast, a move action updates the agent's position without depositing material. Agents do not retain memory of previously observed regions and have no access to global information about either the boundary conditions or the structure being constructed, unless that information is entirely contained within the agent's radius of vision.

\section{Results}

Using the framework described above, we examine how local deposition rules can direct three global geometric properties of networks built from line-segments as building blocks. These include the total area ratio covered by the network, $A_{\rm r}$, the average line density within the covered region, $\overline{\rho}$, and the curvature of the advancing front, $K$. These properties are not exhaustive, but they provide a representative set of targets for demonstrating the procedure shown in Fig.~\ref{fig:rule_design}. Further, because these observables are coupled, they also allow us to test whether rule-sets and their parameters can be selected to direct one structural property via another property we already can reliably control. To quantify sensitivity to initial conditions, each quantity is evaluated over an ensemble of $100$ independent simulations with randomly assigned initial agent positions. This statistical analysis allows us to measure both the mean emergent response and its variability, and therefore to assess how reliably each property can be directed by a given rule-set.

\subsection{Area Coverage Ratio}

We first consider how local deposition rules can direct the area covered by the resulting networks. For this property, the observable $Q$ in Eq.~\ref{eq:expected-value} is the area coverage ratio, $A_{\rm r}$. Because the deposited line segments are one-dimensional objects, we assign each segment an effective thickness $h=10^{-2}L$, where $L$ is the characteristic length of the global domain, defined here as the length of the scaffold branches. We then estimate the covered area by discretizing the global domain into square pixels of side length $\ell_{\rm p}=2h$, and classifying a pixel as covered if any portion of a deposited line segment passes through it. If $N_{\rm p}$ pixels are covered, the covered area is approximated as $A_{\rm cov} \approx N_{\rm p}\ell_{\rm p}^2$. The total available area, $A_{\rm tot}$, is defined as the area of the convex hull of the boundary geometry. The covered area ratio $A_{\rm r}=A_{\rm cov}/A_{\rm tot}$ is then computed at each time step and averaged over an ensemble of $100$ simulations initialized with randomized agent positions.

\textit{Stochastic Baseline}. The area coverage ratio increases only when an agent lays down a line segment that reaches areas that were not covered before. In the notation of Eq.~\ref{eq:expected-value}, the expected increase depends on both the probability that a segment deposited at time $t$ covers previously uncovered area, $p_{\rm t}$, and on the resulting increase in the covered area ratio, $(\delta A_{\rm r})_{\rm t}$. The stochastic rule-set does not deliberately change either quantity. The agent chooses randomly whether to build or move, so it does not preferentially build at times when new area can be covered. When it does build, it chooses its endpoint uniformly at random from the locally available points, so it does not bias segment placement toward uncovered regions. The stochastic baseline therefore provides a model in which both the occurrence and the magnitude of increases in area coverage arise from unbiased random sampling, rather than from systematic changes to either $p_{\rm t}$ or $(\delta A_{\rm r})_{\rm t}$.

\begin{figure*}
    \centering
    \includegraphics[width=1\linewidth]{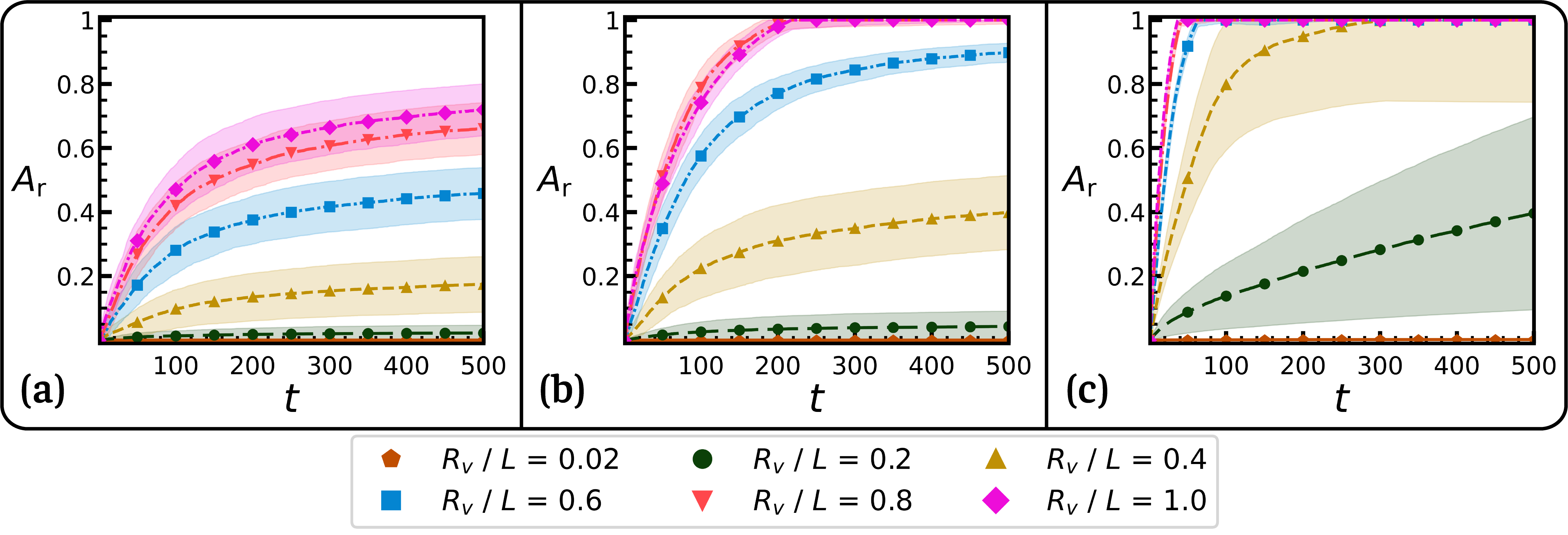}
    \caption{\small{Area coverage ratio, $A_{\rm r}$, as a function of time, $t$, for various values of $R_{\rm v}/L$ with $N_{\rm a}=1$ agent under the (a) stochastic baseline, (b) Max Distance, and (c) Area Threshold rule-sets. Shaded regions indicate one standard deviation across $100$ independent simulations.}}
    \label{fig:directed-area}
\end{figure*}

This interpretation explains the trends in Fig.~\ref{fig:null-area}. Fig.~\ref{fig:null-area}(a) shows a representative random network formed after $10^4$ time steps and $N_{\rm a}=1$ agent. Even after this long construction time, only a small fraction of the available area is covered. This slow growth is shown quantitatively in Fig.~\ref{fig:null-area}(b), where the covered area ratio initially increases rapidly, but then approaches a saturated state. The curves are well described by the form
\begin{equation}
A_{\rm r}(t)=\frac{A_{\rm max} \ t}{\tau+t},
\label{eq:saturation}
\end{equation}
where $A_{\rm max}$ is the area coverage ratio where asymptotic behavior occurs and $\tau$ is the time required to reach half of this value. The black solid lines in Fig.~\ref{fig:null-area}(b) show fits of Eq.~\ref{eq:saturation} to the simulation data. Increasing $R_{\rm v}$ raises the asymptotic value $A_{\rm max}$, but this value remains below full coverage for all possible radii of vision. We note that for $R_{\rm v}/L\geq 0.5$, the agent can access nearly the entire environment from many locations and therefore no longer relies on strictly local information. Even for $R_{\rm v}=L$, however, the asymptotic area coverage remains limited to $A_{\rm max}\approx 0.85$. This saturation occurs because as the network grows, a randomly deposited segment is increasingly likely to pass through regions that are already covered. Consequently, the probability that a new segment covers previously uncovered area decreases over time, causing the growth of $A_{\rm r}$ to slow significantly.

Increasing the number of agents, $N_{\rm a}$, produces an effect similar to increasing the construction time $t$. For a fixed radius of vision and a fixed number of time steps, additional agents increase the accumulated coverage by increasing the number of random trials performed in parallel. This behavior is reflected in Fig.~\ref{fig:null-area}(c), where $A_{\rm r}$ is measured after $t=100$ time steps and eventually plateaus as $N_{\rm a}$ increases. The black solid lines again show fits of Eq.~\ref{eq:saturation} to the simulation data, indicating that the dependence of $A_{\rm r}$ on $N_{\rm a}$ exhibits the same asymptotic form as its dependence on $t$. This is because agents in the stochastic baseline build independently rather than through coordinated construction, so increasing $N_{\rm a}$ effectively adds parallel random trials without introducing a state-dependent bias toward uncovered regions.

The slow growth of $A_{\rm r}$ in the stochastic baseline, particularly when agents have access only to local information $(R_{\rm v}/L<0.5)$, limits the range of area coverage values that can be reached with this rule-set. Although increasing $R_{\rm v}$, $N_{\rm a}$, or $t$ raises the mean value of $A_{\rm r}$, the increase saturates and, on average, does not provide access to the full available range. The standard deviation across $100$ realizations is also large for most values of $R_{\rm v}$, as shown with shaded regions in Figs.~\ref{fig:null-area}(b) and (c), indicating that the stochastic rule-set does not reliably yield reproducible area coverage ratios. Thus, the baseline falls into the regime of Fig.~\ref{fig:rule_design} where the rule-set itself must be modified in order to access target values across the range $0<A_{\rm r}\leq 1$ with low variability.

\textit{Rule-Set Modification}. To design rule-sets that can access and reliably direct area coverage ratios beyond what is achievable with the stochastic baseline, Eq.~\ref{eq:expected-value} suggests two possible mechanisms. A rule can systematically increase the probability $p_{\rm t}$ that during time step $t$, an agent deposits a segment that covers previously uncovered areas. It can also increase $(\delta A_{\rm r})_{\rm t}$, the corresponding increase in the covered area ratio, or do both. Here, we introduce two directed rule-sets that modify the stochastic baseline in increasingly targeted ways to demonstrate how this can be achieved.

The first rule-set, which we call the Max Distance rule-set, is the simplest modification of the stochastic baseline because it changes only the build rule. Instead of selecting an endpoint uniformly at random from all available points within the radius of vision, an agent builds to a point that is maximally distant from its current position while remaining within $R_{\rm v}$. If more than one point satisfies this condition, one is selected uniformly at random. The decision to build or move, as well as the movement rule, remain fully stochastic. This rule-set tests whether increasing the typical length of deposited segments is sufficient to increase $(\delta A_{\rm r})_{\rm t}$ and thereby expand the accessible range of area coverage.

The second rule-set, which we call the Area Threshold rule-set, uses local measurements of area coverage to bias both the decision rule and the build rule, while leaving the movement rule stochastic. The agents are assigned an area coverage ratio threshold $A_{\rm T}$. At each time step, the agent estimates the local area coverage ratio $A_{\ell}$ within its radius of vision by discretizing the region into a Cartesian grid of square cells (see the Supplementary Information for more details). If $A_{\ell}<A_{\rm T}$, the agent considers candidate endpoints within its $R_{\rm v}$ and deposits the segment that produces the largest increase in local area coverage ratio. If $A_{\ell}$ exceeds $A_{\rm T}$, the agent moves to a randomly selected point within its $R_{\rm v}$. In contrast to the Max Distance rule-set, the Area Threshold is designed to increase $p_{\rm t}$ by causing agents to build in locally under-covered regions, and to increase $(\delta A_{\rm r})_{\rm t}$ by selecting the segment that maximizes the local area coverage gain.

Figs.~\ref{fig:directed-area}(b) and (c) show the results of these two modified rule-sets and compare them with the stochastic baseline shown in Fig.~\ref{fig:directed-area}(a). Fig.~\ref{fig:directed-area}(b) depicts the area coverage ratio as a function of time for the Max Distance rule-set for several values of $R_{\rm v}$ with $N_{\rm a}=1$ agent. Restricting the build endpoint to the farthest available points produces only a modest increase in area coverage for $R_{\rm v}/L<0.2$, but leads to a substantial improvement for larger values of $R_{\rm v}$. For sufficiently large $R_{\rm v}$, the Max Distance rule-set makes the full range $0<A_{\rm r}\leq 1$ accessible, while also reducing the ensemble standard deviation relative to the stochastic baseline. However, this performance depends strongly on the radius of vision. Values of $A_{\rm r}>0.80$ are accessible only when $R_{\rm v}/L>0.6$, where the agent can sense a large fraction of the environment and the rule-set is therefore no longer strictly local. In Fig.~\ref{fig:directed-area}(c) we show the results of the Area Threshold rule-set, which throughout this work sets the threshold parameter to $A_{\rm T} =1$. With this rule-set, the full range $0<A_{\rm r} \leq 1$ can be reached with $R_{\rm v}/L<0.5$, meaning that the rule-set can access all target coverage values using strictly local information. However, this increased accessibility comes at the cost of even larger standard deviations than those observed in the stochastic baseline.

\begin{figure}[h]
    \centering
    \includegraphics[width=.85\linewidth]{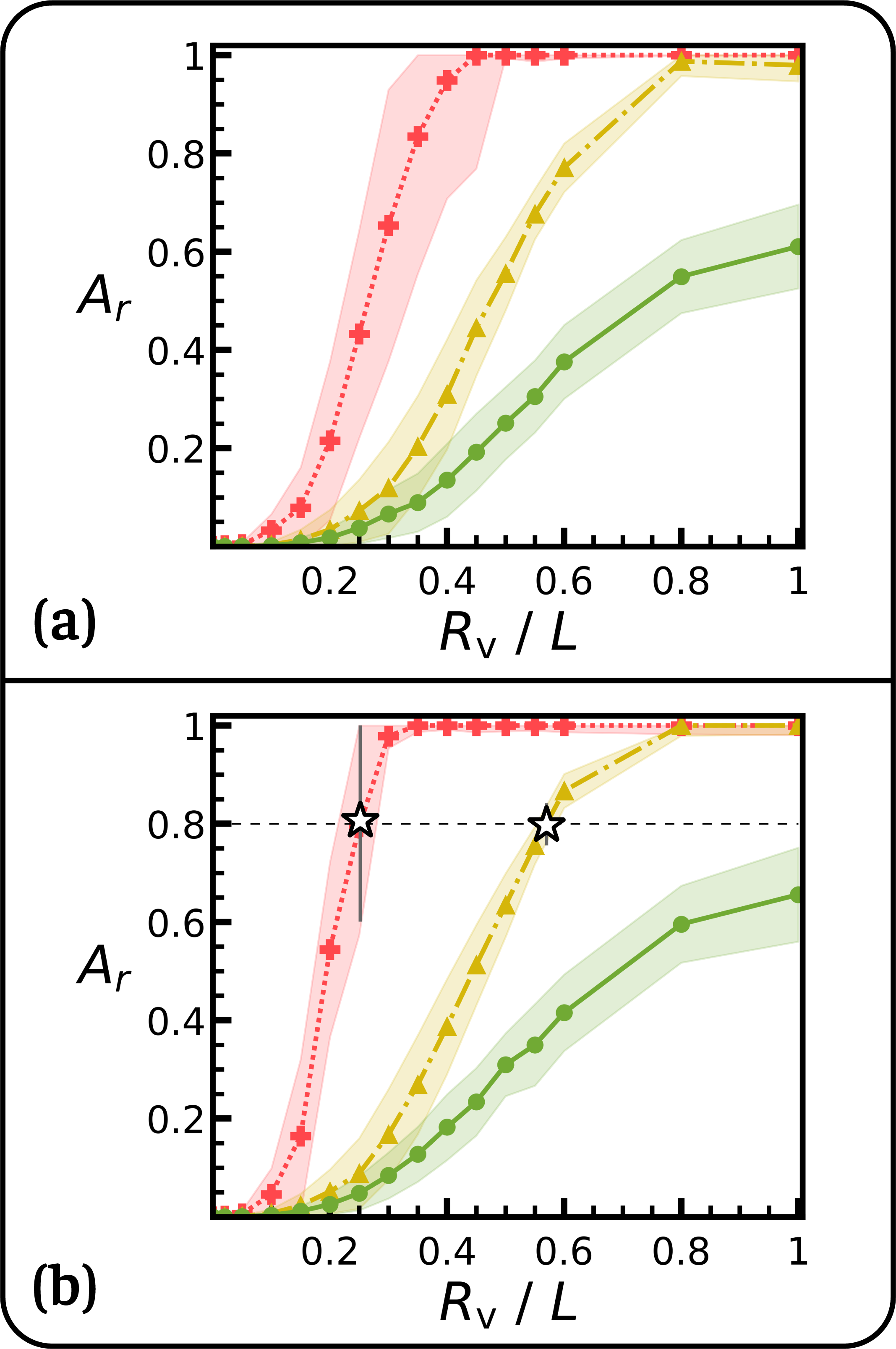}
    \caption{\small{Area coverage ratio, $A_{\rm r}$, as a function of $R_{\rm v}/L$ after $t=200$ time steps for (a) $N_{\rm a}=1$ agent, and (b) $N_{\rm a}=2$ agents. Green circles, yellow triangles, and red plus markers show the stochastic baseline, Max Distance, and Area Threshold rule-sets, respectively. The horizontal dashed line in (b) indicates the target value, and black stars mark directed area values at $A_{\rm r}=0.807, 0.799$  for Area Threshold and Max Distance respectively with a maximum $\epsilon = 0.007$. Shaded regions show one standard deviation across $100$ independent simulations.}}
    \label{fig:all-area}
\end{figure}

\begin{figure*}
    \centering
    \includegraphics[width=1\linewidth]{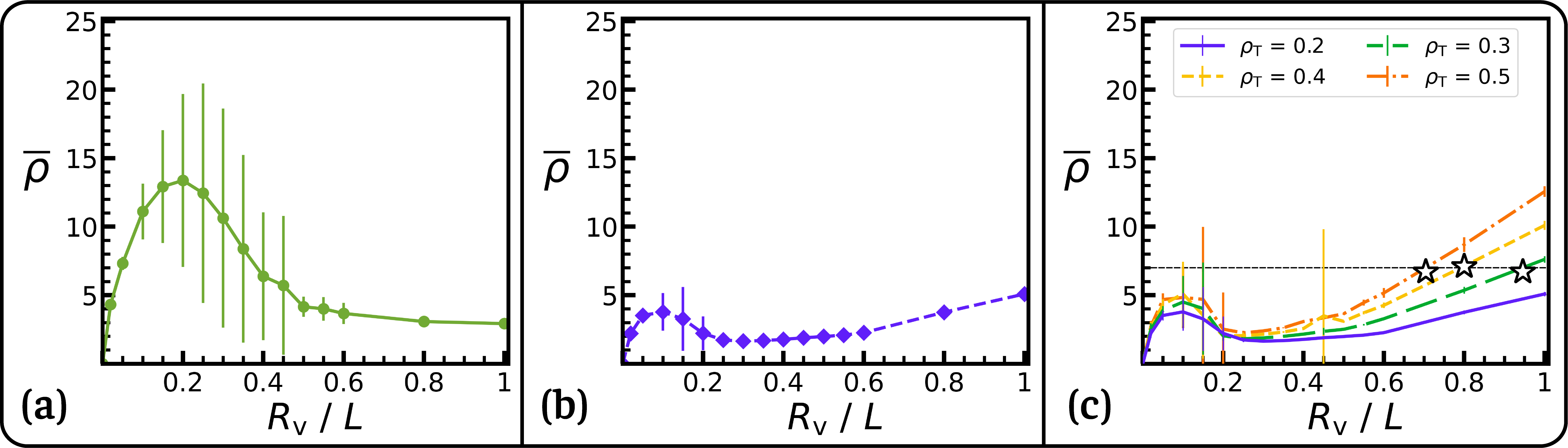}
    \caption{\small{Mean line density, $\overline{\rho}$, as a function of $R_{\rm v}/L$ over $t=500$ time steps with $N_{\rm a}=2$ agents for the (a) stochastic baseline, (b) Density Threshold rule-set with $\rho_{\rm T}=0.2$, and (c) Density Threshold rule-set for various values of $\rho_{\rm T}$. The horizontal dashed line in (c) indicates the target value, and black stars mark directed $\overline{\rho}$ values at $\overline{\rho} = 6.7, 6.7, 7.1$ with a maximum $\epsilon = 0.3$ .}}
    \label{fig:density_directing}
\end{figure*}

To better compare the performance of these rule-sets, we plot the area coverage ratio for the stochastic baseline, Max Distance, and Area Threshold rule-sets as a function of the normalized radius of vision, $R_{\rm v}/L$. In Figs.~\ref{fig:all-area}(a) and (b) the results are shown for simulations with $N_{\rm a}=1$ and $N_{\rm a}=2$ agents, respectively. The Area Threshold has a relatively large standard deviation, $0\leq \sigma \leq 0.34$, when only one agent is used. This occurs because the rule-set drives $A_{\rm r}$ from low coverage to nearly full coverage very quickly, so at a fixed observation time, small differences in the agent's path from one realization to another can produce large differences in the measured area coverage. In one realization, the agent may reach high coverage early, while in another, it may be delayed by a different sequence of local movements. With two agents, the result is less sensitive to any one trajectory. The two agents sample the environment in parallel, so the growth rate is effectively averaged over the paths of two builders, and the spread in $A_{\rm r}$ values across realizations is reduced to $0\leq \sigma \leq 0.24$. For this reason, in what follows, we will use $N_{\rm a} = 2$ agents in our simulations.

\textit{Parameter Selection}. The curves in Fig.~\ref{fig:all-area}(b) can be used as calibration curves for selecting rule parameters that realize a prescribed target value of the area coverage ratio. Suppose, for example, that the desired mean area coverage ratio is $A_{\rm r}^{*}=0.80$ with $\epsilon \leq 0.01$ and $\sigma^{*} \leq 0.25$. The stochastic baseline cannot access this value, even when $R_{\rm v}/L=1$. Both directed rule-sets can reach this target, although the Max Distance rule-set requires $R_{\rm v}/L>0.5$ and therefore access to global information. For each directed rule-set, we interpolate the corresponding $A_{\rm r}(R_{\rm v}/L)$ curve to identify the value of $R_{\rm v}/L$ expected to produce $A_{\rm r}^{*}=0.80$. We then run $100$ parallel simulations using these interpolated parameter values and verify that the resulting networks reliably achieve the prescribed mean area coverage ratio, as indicated by the star markers in Fig.~\ref{fig:all-area}(b). This indicates that while the rule-set modification makes the full range of $A_{\rm r}$ values accessible, the parameter selection step is necessary for identifying the operating point that realizes a specified target within that range. 

\subsection{Mean Network Line Density}

Next we turn our attention to mean network line density, an important property which influences porosity, transport pathways, and the amount of material required for the network construction. Here we demonstrate that the framework shown in Fig.~\ref{fig:rule_design} can be used to direct the mean line density of the agent-built networks toward a target value. We define the local line density by using the same grid discretization introduced for the area coverage ratio and counting the number of deposited line segments that pass through each pixel. For each network, we average this local line density over all non-zero pixels to obtain a single mean density value for the network. We then average this quantity over $100$ independent realizations to obtain the ensemble-averaged mean line density, $\overline{\rho}$.

\textit{Stochastic Baseline}. We first evaluate the stochastic baseline using the same fully random rule-set introduced for the area coverage ratio. Fig.~\ref{fig:density_directing}(a) shows $\overline{\rho}$ as a function of the normalized radius of vision, $R_{\rm v}/L$ for networks built with $N_{\rm a} = 2$ agents in $t=500$ time steps. For small $R_{\rm v}/L$, the stochastic baseline produces a relatively wide range of $\overline{\rho}$ values. However, this range is accompanied by large standard deviations ($0.2 \leq \sigma \leq 8.0$), especially when deposition is strictly local $(R_{\rm v}/L<0.5)$. Thus, although different density values can be accessed, they are not produced reliably across realizations. For larger $R_{\rm v}/L$, the variability decreases, but the accessible range of $\overline{\rho}$ becomes comparatively narrow. The stochastic baseline therefore provides limited control over the average line density since depending on $R_{\rm v}/L$, it either accesses a broad but noisy range of densities or a narrow but more reproducible range.

\textit{Rule-Set Modification}. To direct the mean line density toward prescribed target values, we introduce a Density Threshold rule-set that uses local density measurements to decide whether an agent should build or move. Unlike the area coverage ratio, the mean line density does not have a fixed upper bound. We therefore do not aim to span a predetermined interval such as $[0-1]$. Instead, the objective is to show that a directed rule-set can systematically shift the accessible density values and do so with lower variability than the stochastic baseline.

In this rule-set, each agent is assigned a local density threshold, $\rho_{\rm T}$. At each time step, the agent computes the local line density $\rho_{\ell}$ within its radius of vision and compares this value to $\rho_{\rm T}$ (see the Supplementary Information for more details). If $\rho_{\ell}<\rho_{\rm T}$, the agent builds. If $\rho_{\ell}\geq \rho_{\rm T}$, the agent moves to a maximally distant point within its radius of vision to explore a new region.  
The local density threshold $\rho_{\rm T}$ is related to the global mean density, $\overline{\rho}$, through the conversion between the density measured within the agent's radius of vision and the density measured on the global pixel grid. Under approximately uniform density, this gives $\overline{\rho}\approx \rho_{\rm T}\ell_0 R_{\rm v}/N_{\rm cc}$, where $\ell_0$ is the global pixel size and $N_{\rm cc}$ sets the grid resolution within the agent's radius of vision. Thus, changing $\rho_{\rm T}$ provides a direct local mechanism for shifting the ensemble-averaged mean line density.

When building, the agent deposits the segment that produces the largest increase in local area coverage, as in the Area Threshold rule-set. This prevents the target line density from being achieved only by repeatedly adding lines to a small region, and instead promotes expansion of the network as density is accumulated. When moving, the agent moves to one of the points that is maximally distant from its current location, allowing it to leave locally dense regions and search more effectively for regions where $\rho_{\ell}<\rho_{\rm T}$. In terms of Eq.~\ref{eq:expected-value}, the decision rule in this rule-set primarily increases the probability $p_{\rm t}$ that an agent deposits lines in a region whose local density remains below the prescribed threshold, while the build and movement rules only help distribute that deposition over the growing network.

Fig.~\ref{fig:density_directing}(b) shows the resulting values of $\overline{\rho}$ for the Density Threshold rule-set with threshold density $\rho_{\rm T}<0.2$ as a function of the normalized radius of vision, $R_{\rm v}/L$. Compared with the stochastic baseline in~\ref{fig:density_directing}(a), the Density Threshold rule-set produces substantially smaller standard deviations ($0.0 \leq \sigma \leq 2.3$), across realizations. This reduction in variability arises because the decision rule promotes more uniform deposition throughout the structure as agents do not continue building in regions that have already exceeded $\rho_{\rm T}$.

For a fixed value of $\rho_{\rm T}$, the range of accessible mean line densities remains relatively narrow. This does not indicate failure of the rule-set. Rather, it shows that the accessible range must be explored by varying the relevant parameters such as $\rho_{\rm T}$. Fig.~\ref{fig:density_directing}(c) demonstrates that changing $\rho_{\rm T}$ systematically shifts the resulting values of $\overline{\rho}$. We report results up to $\rho_{\rm T}\leq 0.5$, because for larger values of $\rho_{\rm T}$, $500$ time steps are insufficient for two agents to reach the imposed local density threshold and maintain relatively low variability. This limitation is practical rather than fundamental, since increasing the construction time or the number of agents should allow access to higher values of $\overline{\rho}$.

\textit{Parameter Selection}. Here, $\rho_{\rm T}$ acts as an additional rule parameter for directing the mean line density toward a target value. For example, suppose the desired mean line density is $\overline{\rho}^{*}=7.0$ lines per unit area with an acceptable standard deviation of $\sigma^* \leq 0.5$ and $\epsilon \leq 0.3$. Using Fig.~\ref{fig:density_directing}(c) as our calibration plot, we can identify the value of $\rho_{\rm T}$ that can reliably produce this density value, interpolate the corresponding $\overline{\rho}(R_{\rm v}/L)$ curve to select the required value of $R_{\rm v}/L$, and then run ensemble simulations using this set of parameters. The star markers in Fig.~\ref{fig:density_directing}(c) show that this procedure successfully predicts the value of $R_{\rm v}/L$ and is able to produce $\overline{\rho}^{*}=7.0 \pm 0.3$ for all three cases of the Density Threshold rule-set when averaged over $100$ realizations with a maximum standard deviation of $\sigma = 0.5$. In this case, the local density threshold is used to select the accessible density regime, and the radius of vision selects the operating point within it.

\subsection{Absolute Total Curvature}

\begin{figure}[h!]
    \centering
    \includegraphics[width=0.85\linewidth]{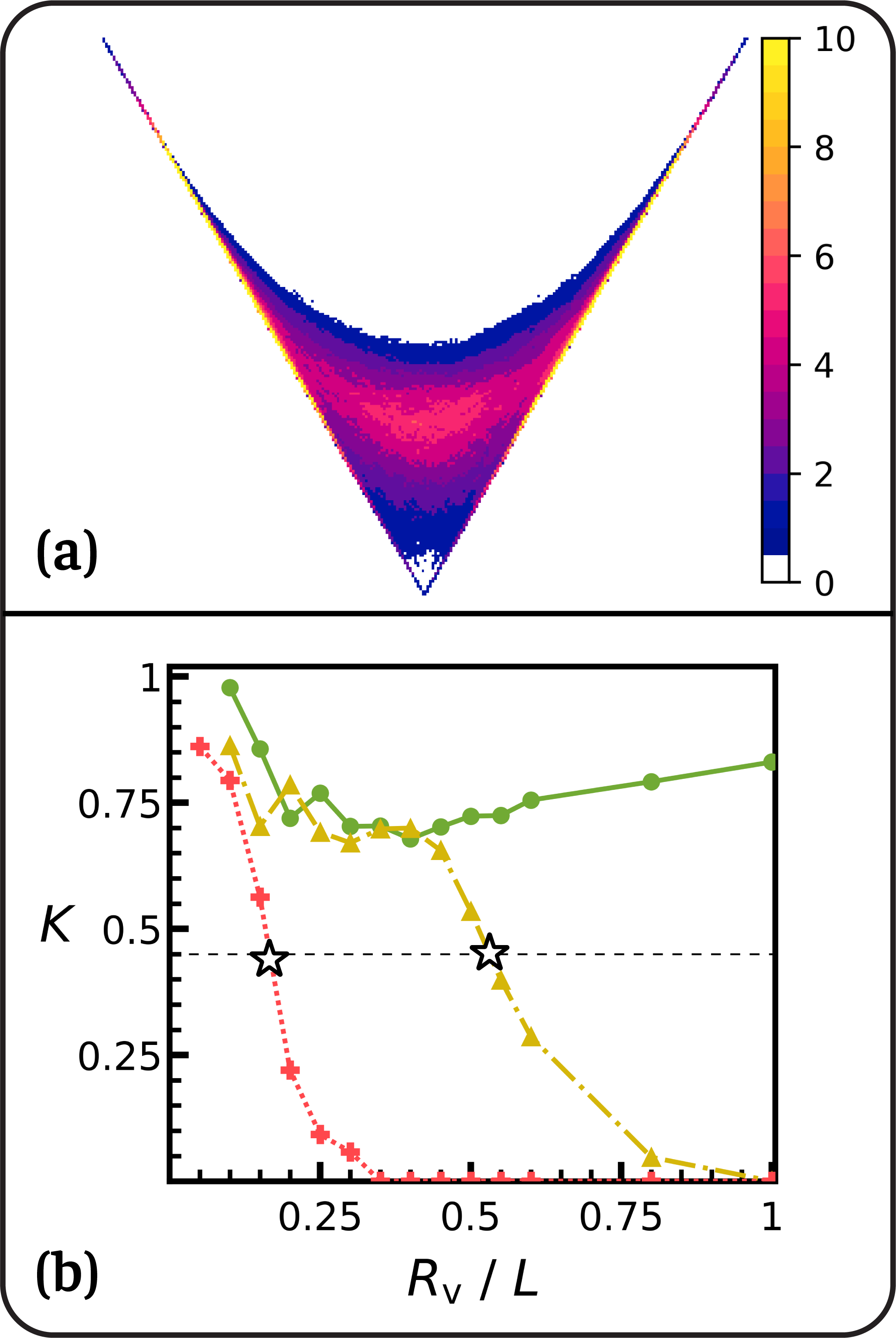}
    \caption{\small{(a) Spatial density field for networks built using the stochastic baseline with $N_{\rm a}=2$ agents, $R_{\rm v}/L=0.5$, and $t=500$ time steps. Each pixel value is ensemble-averaged over $100$ independent simulations. Similar density fields are used to extract front curvature for each value of $R_{\rm v}/L$ and each rule-set. (b) Normalized front curvature, $K$, as a function of $R_{\rm v}/L$, where green circles, yellow triangles, and red plus markers show the stochastic baseline, Max Distance, and Area Threshold rule-sets, respectively. The horizontal dashed line in (b) indicates the target value, and black stars mark directed values of $K = 0.46, 0.45$ for the Area Threshold and Max Distance rule sets respectively with a maximum $\epsilon = 0.01$. }}
    \label{fig:curvature}
\end{figure}

We next consider the curvature of the advancing network front, a global geometric property that can be directed through local deposition rules. We define the network \textit{front} as the maximum vertical position reached by the network at each horizontal position. This quantity is of particular interest because curved fronts resemble the curved surfaces observed in caterpillar tents, as shown in Fig.~\ref{fig:hexagonal_images1}, which lead us to question how fundamental this property is to network structures built within branched boundary conditions.  

We observe that the emergence of front curvature in networks built by line deposition is strongly influenced by the boundary geometry. In the systems studied here, line segments are deposited within a scaffold formed by two branches meeting at a convex angle $\theta=\pi/6$. This scaffold constrains the accessible orientations of deposited segments and imposes a position dependent upper bound on the height that the network can reach. Near the scaffold boundaries, deposited segments tend to align with the angled branches, whereas near the center of the scaffold they are deposited at smaller angles relative to the horizontal direction. This spatial variation in segment orientation shapes the free boundary of the network envelope and gives rise to an emergent front curvature.

Here, we extract the curvature from the ensemble-averaged density field rather than from individual networks. This averaging suppresses front roughness associated with individual realizations and reveals the global geometry of the emerging networks as shown in Fig.~\ref{fig:curvature}(a). From this ensemble-averaged field, we identify the front and fit a smooth curve $f(s)$ to the resulting profile, where $s$ denotes arc length along the front. We then compute the normalized absolute total curvature,
\begin{equation}
K= \frac{1}{\pi}\int |\kappa(s)|ds ,
\label{eqn:abs_tot_curve}
\end{equation}
where $\kappa(s)=f^{\prime\prime}(s)/ \left[1+f^{\prime}(s)^2\right]^{3/2}$. The normalization by $\pi$ gives $0 \leq K < 1$, where $K=0$ corresponds to a flat front and larger values correspond to more strongly curved fronts.

\textit{Stochastic Baseline}. We first evaluate the curvature produced by the stochastic baseline, using the same fully random rule-set introduced for the area coverage ratio and mean line density. Data marked with green circles in Fig.~\ref{fig:curvature}(b) show the resulting values of $K$ as a function of the normalized radius of vision, $R_{\rm v}/L$ for networks built with $N_{\rm a} =2$ agents in $t=500$ time steps. The stochastic baseline produces curvature values only within a relatively narrow interval, approximately $0.75 \lesssim K < 1$. In particular, $K$ remains close to $0.75$ over a broad range of $R_{\rm v}/L$, indicating that in this case varying the radius of vision alone is not sufficient to access weakly curved or flat fronts. This restricted range is a direct consequence of the saturation of the area coverage ratio in the stochastic baseline where $A_{\rm r}$ grows initially but eventually saturates below full coverage as was demonstrated in Fig.~\ref{fig:null-area}. This incomplete filling preserves the curved envelope imposed by the scaffold geometry. A flat front requires the network to cover the available area up to the upper boundary across the horizontal extent of the scaffold. In this limit, $A_{\rm r}\rightarrow 1$ and the front curvature approaches $K\rightarrow 0$. Because the stochastic baseline cannot reliably drive $A_{\rm r}$ to $1$, it also cannot access the full range $0 \leq K <1$. Thus, the baseline again falls into the regime of Fig.~\ref{fig:rule_design} where the rule-set must be modified.

\textit{Rule-Set Modification}. The coupling between front curvature and area coverage suggests that rule-sets designed to increase $A_{\rm r}$ can also be used to access lower values of the front curvature. We therefore test whether the Max Distance and Area Threshold rule-sets, introduced earlier for directing area coverage ratio, can also serve as mechanisms for directing the absolute total curvature of the front.
Because these two rule-sets expand the accessible range of $A_{\rm r}$, they can also expand the accessible range of $K$. 
Fig.~\ref{fig:curvature}(b) shows that both rule-sets produce values of $K$ below those accessible to the stochastic baseline. In particular, by promoting deposition into previously uncovered regions and pushing $A_{\rm r}$ to $1$, they make the regime $0 \leq K < 0.75$ accessible. Since the Area Threshold rule-set actively maximizes area coverage at each step, it leads to the formation of networks with defined curvature at smaller $R_{\rm v}/L$ values and therefore appears earlier in Fig.~\ref{fig:curvature}(b). We note that results shown in this plot do not imply that $A_{\rm r}$ and $K$ are independently prescribed at the same time. Rather, curvature is the target observable in this scenario, and the area coverage rule-sets provide a local mechanism for shifting the network into curvature regimes that are inaccessible to the stochastic baseline. Thus, $K$ is directed indirectly by exploiting its geometric coupling to area coverage.

\textit{Parameter Selection}. Once the directed rule-sets make the desired curvature range accessible, the requisite radius of vision can be selected to realize a prescribed target value of $K$. For example, suppose the desired absolute total curvature is $K^{*}=0.45$ with $\epsilon \leq 0.02$. We use the calibration curves $K(R_{\rm v}/L)$ in Fig.~\ref{fig:curvature}(b) for the Max Distance and Area Threshold rule-sets, interpolate each curve to identify the relevant value of $R_{\rm v}/L$ expected to produce $K^{*}$, and then run independent ensemble simulations using these parameter values. The results shown by the star markers in Fig.~\ref{fig:curvature}(b) verify that the selected parameters produce the target curvature.

Front curvature provides an example in which one global property can be directed by exploiting its coupling to another. Rather than designing an entirely new rule-set, we use the previously developed area coverage rule-sets to tune the large scale front geometry of the resulting networks. This illustrates how the rule design framework can be applied not only to independent observables, but also to strongly coupled properties where directing one quantity provides a mechanism for directing another.

\section{Conclusion}

In summary, we have explored how to systematically design behavioral rules for simple agents such that, when they follow these rules to manipulate and deposit building blocks locally, they construct structures whose global properties emerge in a controlled way. These properties arise not only from how building blocks connect to one another, but also from how agents interact with the evolving structure, their surrounding environment, and the decisions they make during deposition. Throughout this work, we have shown that these local behavioral rules can be designed and refined so that emergent properties of network structures, such as area coverage, mean line density, and front curvature, can be reliably directed toward desired target values.

What we have studied here represents only a small fraction of what is possible. In this work, we have primarily focused on designing rules for individual agents and considered only a minimal form of group behavior, in which the number of agents increases without introducing interactions among them.
Yet it is in the presence of collective that the problem becomes richer. This raises the question of how interactions among agents should be designed so that agents do not simply build alongside one another, but instead build together. It also points to the need to understand what kinds of structures become possible when agents share information, coordinate their actions, or adapt to one another in real time. There remains a wide and largely unexplored space in understanding how local interactions between agents, and not only between agents and their environment, shape emergent structure. More broadly, many fundamental questions remain open. For instance, the mapping between available degrees of freedom, the rules used to tune them, and the structures that ultimately emerge is still poorly understood. It remains unclear what minimal set of degrees of freedom is required to reliably produce a desired emergent property, and how much information about the environment is necessary to direct emergence.

We will pursue these directions in future work. The results presented here point toward an alternative way of thinking about how structures may be designed and built in environments where centralized control or continuous human supervision is not possible. Rather than exerting top-down specification over outcomes, the goal in such settings would be to define proper constraints that guide the self-organization of simple agents toward building functional structures. We view this as a step forward in extending human capabilities for engineering in unfamiliar and uncertain environments.

\section{Acknowledgments}
We would like to thank Avaneesh Narla, Suraj Shankar, Randy Ewoldt, Sameh Tawfick, and William King for several inspiring discussions that helped shape how we thought about this problem. We also thank Sascha Hilgenfeldt and Francesco Zamponi for early feedback on the manuscript.
We are grateful to the Jasper Ridge Biological Preserve at Stanford University for opening its grounds to us. We are especially grateful to Avaneesh Narla for his patience, persistence, and sharp eye in helping us find such small creatures (tent caterpillars) across a very large landscape. Without his help, we would not have been able to make the observations that started this project.
This work was supported by start-up funds from the University of Illinois, which we gratefully acknowledge.

\bibliography{Refs}

\clearpage

\section{Supplemental Information}

\subsection{Local Area Measure}

\begin{figure}[hpt!]
    \centering
    \includegraphics[width=.712\linewidth]{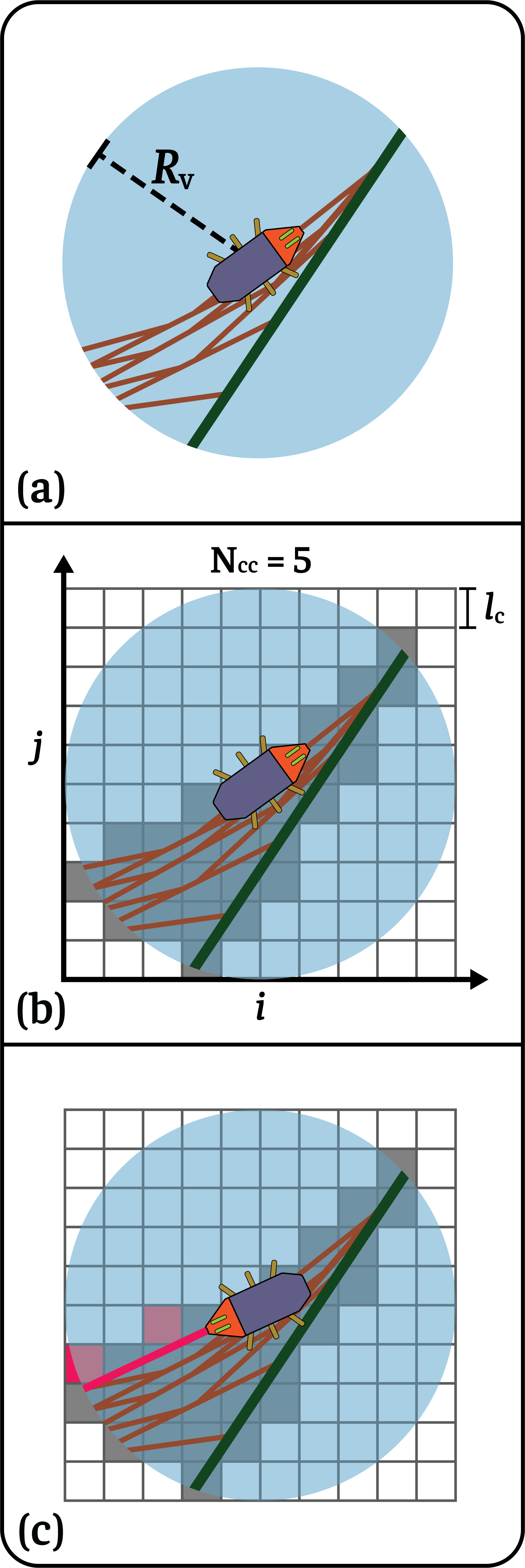}
    \caption{(a) The agent begins by discretizing the lines in the environment within its radius of vision into a set of available points. These points are not shown to avoid making the figure overly dense. (b) The surrounding domain is then discretized into a Cartesian grid, and the agent measures the current local area coverage ratio by evaluating the number of cells within the grid that contain one or more points. (c) Finally, the agent evaluates all candidate segments using this same measurement method and deposits the line segment that maximizes the increase in the local area coverage ratio.}
    \label{fig:area_target_algo}
\end{figure}

In the Area Threshold rule-set, a measure of the local area coverage must be derived from information already available to agents, such that they can determine which segment will locally maximize the increase in area coverage. Recalling from Fig.~\ref{fig:agent_construction} that agents perceive their surroundings as a discrete set of available points, we discretize the domain within their radius of vision into grid cells and determine how many cells contain at least one available point. In $2$D, we consider this discretization to be a uniform Cartesian grid, which we represent as a matrix $A_{\rm ij}$, where we record a $1$ at each pair of indices for which one or more points are present. By doing so, we estimate the covered area by assuming that if at least one point lies within a cell, then there must be a line segment that passes through it, and thus the cell is covered.
Here we set the number of cells along each cardinal direction from the center of the mesh, $N_{\rm cc}=50$, and allow the characteristic length of each cell, $\ell_{\rm c}$, to vary with respect to $R_{\rm v}$, such that $\ell_{\rm c} = R_{\rm v}/N_{\rm cc}$. With this information, we compute a local measure of the average ratio of covered area, $A_{\rm \ell}$, according to Eq.~\ref{eqn:relative_area},   

\begin{equation}
    A_{\rm \ell} = \frac{l_{\rm c}^2\sum\limits_{i = 1}^{2N_{\rm cc}}\sum\limits_{j = 1}^{2N_{\rm cc}}A_{\rm ij}}{\pi R_{\rm v}^2}.
    \label{eqn:relative_area}
\end{equation} 
In Fig.~\ref{fig:area_target_algo} we display an example of how this local measure is used in the ARA rule-set to evaluate optimal line segments to maximize local area coverage.

\subsection{Local Density Measure}

In the Density Threshold rule-set, we similarly require a measure of the local density and thus apply the same discretization scheme to the available points as described in the Local Area Measure section above. Here, however, instead of only counting whether at least one point is contained within a cell, we count the total number of points that belong to distinct lines within each cell and record this value in the matrix $D_{\rm ij}$. From this, we can then compute a measure of the average local density using Eq.~\ref{eqn:relative_density},   

\begin{equation}
    \rho_{\rm \ell} = \frac{\sum\limits_{i = 1}^{2N_{\rm cc}}\sum\limits_{j = 1}^{2N_{\rm cc}}D_{\rm ij}}{\pi R_{\rm v}^2}.
    \label{eqn:relative_density}
\end{equation}
\newpage

\end{document}